\documentclass[prd,twocolumn,amsmath,amssymb,nofootinbib,superscriptaddress]{revtex4-1}
\usepackage{array,mathtools,amssymb,booktabs,multirow,diagbox,hhline,natbib}
\pdfoutput=1

\usepackage{graphicx,graphics,color}
\usepackage{dcolumn}
\usepackage{bm}
\usepackage{bbold}
\usepackage{amsmath}
\usepackage{xcolor}

\allowdisplaybreaks

\newcommand{\Od}{{\cal O}}

\newcommand{\qint}{\int \frac{d^3 \vec{q}}{(2\pi)^3}}
\newcommand{\im}{\mbox{Im}\,}

\newcommand{\intT}{\int_0^{1/T} d\tau \int d^3 \vec{x}}

\newcommand{\mean}[1]{\left\langle{#1}\right\rangle}

\newcommand{\conds}{\langle \bar s s \rangle}
\newcommand{\condl}{\mean{\bar q q}}

\newcommand{\diff}{\text{d}}
\newcommand{\gsim}{\raise.3ex\hbox{$>$\kern-.75em\lower1ex\hbox{$\sim$}}}

\begin{document}

\title{The role of strangeness in chiral and $U(1)_A$ restoration}
\author{A. G\'omez Nicola}
\email{gomez@ucm.es}
\affiliation{Departamento de F\'{\i}sica
Te\'orica and IPARCOS. Univ. Complutense. 28040 Madrid. Spain}
\author{J. Ruiz de Elvira}
\email{elvira@itp.unibe.ch}
\affiliation{Albert Einstein Center for Fundamental Physics, Institute for Theoretical Physics,
University of Bern, Sidlerstrasse 5, CH--3012 Bern, Switzerland}
\author{A. Vioque-Rodr\'iguez}
\email{avioque@ucm.es}
\affiliation{Departamento de F\'{\i}sica
Te\'orica and IPARCOS. Univ. Complutense. 28040 Madrid. Spain}
\author{D. \'Alvarez-Herrero}
\email{davial06@ucm.es}
\affiliation{Departamento de F\'{\i}sica
Te\'orica and IPARCOS. Univ. Complutense. 28040 Madrid. Spain}

\begin{abstract}
  We use recently derived Ward identities and lattice data for the light- and strange-quark condensates to reconstruct the scalar and pseudoscalar susceptibilities ($\chi_S^\kappa$, $\chi_P^K$) in the isospin 1/2 channel.  We show that $\chi_S^\kappa$ develops a maximum above the QCD chiral transition, after which it degenerates with $\chi_P^K$.  We also obtain $\chi_S^\kappa$ within Unitarized Chiral Perturbation Theory (UChPT) at finite temperature, when it is saturated with the $K_0^*(700)$ (or $\kappa$) meson, the dominant lowest-energy state in the isospin 1/2 scalar channel of $\pi K$ scattering.  Such UChPT result reproduces the expected peak structure, revealing the importance  of thermal interactions, and makes it  possible to examine the $\chi_S^\kappa$ dependence on the light- and strange-quark masses.  A consistent picture emerges controlled by the $m_l/m_s$ ratio that allows one studying 
$K-\kappa$ degeneration in the chiral, two-flavor and $SU(3)$ limits.
These results provide an alternative sign for $O(4)\times U(1)_A$ restoration that can be explored in lattice simulations and highlight the role of strangeness, which regulated by the strange-quark condensate helps to reconcile the current tension among lattice results regarding $U(1)_A$ restoration.

 \end{abstract}

 \maketitle

\section{Introduction}

Chiral symmetry restoration is a key ingredient to understand the QCD phase diagram~\cite{Ratti:2018ksb,Bazavov:2019lgz,Nicola:2020iyl}.  The remarkable advances achieved in lattice simulations have revealed a crossover transition at $T_c\simeq$ 155 MeV for physical quark masses and vanishing chemical potentials ~\cite{Aoki:2009sc,Borsanyi:2010bp,Bazavov:2011nk,Bazavov:2014pvz,Bazavov:2018mes}. In the light  chiral limit $m_l\rightarrow 0$  this pseudocritical behavior is expected to become a ``true'' phase transition with a critical temperature $T_c^0\simeq$ 132 MeV~\cite{Ding:2019prx}. 
Nevertheless, the universality class and even the order of this transition are still not fully understood and depend on the strength of the breaking of the anomalous $U(1)_A$ symmetry at the critical temperature~\cite{Pisarski:1983ms,Shuryak:1993ee,Pelissetto:2013hqa}.

Thus, the very nature of the chiral transition is intimately connected to $U(1)_A$ restoration. However, there is currently no agreement as to whether this symmetry is effectively restored close to the critical temperature. While several phenomenological~\cite{Cohen:1996ng,Meggiolaro:2013swa,Azcoiti:2016zbi,GomezNicola:2017bhm,Nicola:2018vug} and  lattice~\cite{Aoki:2012yj,Cossu:2013uua,Tomiya:2016jwr,Brandt:2016daq,Brandt:2019ksy} analyses for $N_f=2$  light flavors of mass $m_l=m_u=m_d$ support the idea that the $U(1)_A$ symmetry can be effectively restored at the chiral transition in the chiral limit, lattice results for $N_f=2+1$ (i.e., including the strange quark flavor  with mass $m_s\gg m_l$)  suggest that the anomalous $U(1)_A$ symmetry is still broken in the chiral crossover region~\cite{Buchoff:2013nra,Dick:2015twa,Kaczmarek:2020sif,Ding:2020xlj}. 
This phenomenon  has also implications for the hadron spectrum~\cite{Shuryak:1993ee,Lee:1996zy,Meggiolaro:2013swa} as well as phenomenological effects driven by the associated reduction of the anomalous $\eta'$ mass and the topological susceptibility in a thermal environment~\cite{Kapusta:1995ww,Csorgo:2009pa,Nicola:2019ohb,Lombardo:2020bvn}. 

The main observables commonly employed to study chiral symmetry restoration, both in the lattice and in phenomenological analyses,
are the light-quark condensate $\condl$ and the scalar susceptibility $\chi_S$, being the chiral transition signaled by both the inflection point of  $\condl$ and the peak of $\chi_S$.
Nevertheless, since the axial anomaly is consequence of a short-distance quantum effect, restored only asymptotically~\cite{Gross:1980br},
there is no corresponding order parameter to study $U(1)_A$ restoration.

The manifestation and restoration of global symmetries such as the chiral $SU(2)_L\times SU(2)_R\approx O(4)$ and  the $U(1)_A$ ones can also be studied by analyzing their effect on the temperature-dependent properties of the particle spectrum. 
For instance, states of opposite parity related under axial $SU(2)_A$ rotations---the so-called chiral partners---are expected to degenerate at the chiral transition. 
Correlation functions of chiral partners and properties derived from them (like susceptibilities and screening masses) should also degenerate as the transition is reached.
In the same way, an effective $U(1)_A$ restoration should be indicated by the degeneracy of correlation functions belonging to a $U(2)\times U(2)$ universality class.

So far, available lattice studies looking for the interplay of chiral and $U(1)_A$ restoration concern the isoscalar and isovector channels.
Namely, defining the lightest scalar and pseudoscalar operators in the isospin $I=0,1$ sector as
\begin{align}
 \pi^a&=i\bar q\gamma^5\tau^a q,&\delta^a&=\bar q\tau^a q,\notag\\
 \sigma&=\bar q q,& \eta_l&=i\bar q\gamma^5 q,
\end{align}
with $q$ the light-quark doublet and $\tau$ the Pauli matrices,
chiral symmetry restoration implies $\pi^a-\sigma$ and $\delta^a-\eta_l$ degeneration, where the $\pi^a$, $\sigma$, $\delta^a$ and $\eta_l$  quark bilinears 
correspond to the pion, $f_0(500)$ (or $\sigma$) and the light components of the $a_0(980)$ and $\eta$~\cite{Zyla:2020zbs}, respectively.
Likewise, $\pi^a-\delta^a$  and $\sigma-\eta_l$  are expected to degenerate once the $U(1)_A$ symmetry is effectively restored.
Whereas chiral degeneration around and above $T_c$ has been clearly observed in the lattice both using screening masses~\cite{Cheng:2010fe,Brandt:2016daq,Bazavov:2019www} and susceptibilities~\cite{Buchoff:2013nra}, hence confirming  theoretical predictions at finite temperature and/or density~\cite{Hatsuda:1986gu,Bernard:1987im,Rapp:1999ej,Nicola:2013vma,Heller:2015box,Jung:2016yxl,Ishii:2016dln,Nicola:2018vug}, lattice results concerning $U(1)_A$ restoration are not conclusive;
while $N_f=2$ simulations suggest $\pi-\delta$ degeneration close to the chiral limit~\cite{Aoki:2012yj,Cossu:2013uua,Tomiya:2016jwr} and for physical quark masses~\cite{Brandt:2016daq,Brandt:2019ksy}, $N_f=2+1$ lattice results report sizable differences between the $\pi$ and $\delta$ susceptibilities in the region where $\pi-\sigma$ degeneration occurs~\cite{Buchoff:2013nra,Dick:2015twa}.

In this work we will present a thorough analysis of an alternative sector; namely, the $I=1/2$ channel involving the kaon $K$ and  $K_0^*(700)$ (or $\kappa$) mesons as the lightest pseudoscalar and scalar states, respectively.  The study of susceptibilities in this sector will provide additional evidences regarding chiral and $U(1)_A$ restoration, which will help to reconcile the apparently conflicting scenarios mentioned above and will highlight the role of the strange quark in a explicit and consistent way. 
On the one hand, our analysis is based on Ward Identities (WIs), which predict the behavior of susceptibilities in a channel where there are currently no lattice results available.
Furthermore, $I=1/2$ WIs would provide a tool to study $O(4)\times U(1)_A$ restoration in terms of quark condensates, well controlled lattice quantities as opposed to those customarily used, such as the $\delta,\,\eta$ or topological susceptibilities, which are considerably more noisy~\cite{Buchoff:2013nra,Bonati:2015vqz,Borsanyi:2016ksw,Lombardo:2020bvn}.
On the other hand, we will show that the main properties of $\chi_S^\kappa$ can be described when it is saturated by the thermal pole of the $K_0^* (700)$ meson, which in turn can be generated in unitarized $\pi K$ scattering at finite temperature. This second approach will shed light on the quark mass dependence and the role of thermal interactions.

With the above motivation in mind, the paper is structured as follows: in section~\ref{sec:wi} we will review the relevant WIs involving the $K$ and $\kappa$ susceptibilities, as well as the main properties regarding chiral and $U(1)_A$ transformations in the $I=1/2$ sector. In section~\ref{sec:chips} we obtain some of our main results regarding the properties of these susceptibilities; namely,  we will show  the existence of a maximum in the  $\kappa$ susceptibility signaling degeneration with the $K$ one, which, as we will see, is consistent with asymptotic $O(4)\times U(1)_A$ restoration in the physical case. 
Our conclusions will be reached both from a direct analysis of lattice data to reconstruct the susceptibilities from the WIs, section~\ref{sec:latt}, and from UChPT, section~\ref{sec:uchpt}, which provides a tool to study their behavior towards the  light chiral and $SU(3)$ limits.  
Finally, the consequences of our results regarding $O(4)\times U(1)_A$ restoration will be further addressed in section~\ref{sec:conseq}, where we will examine the different limits of interest. In particular, within the context of the results obtained here and previous ones from WIs, ChPT and phenomenological analyses, we will present some arguments helping to understand the lattice results obtained in two and three flavors. 

\section{ Ward Identities in the strange sector and $K$-$\kappa$ degeneration}
\label{sec:wi}

A useful set of WIs connecting pseudoscalar and scalar susceptibilities with quark condensates for all isospin channels has been recently derived and analyzed in~\cite{Nicola:2013vma,Nicola:2016jlj,GomezNicola:2017bhm,Nicola:2018vug}. In particular, WIs in the $I=1/2$ sector read
\begin{align}
\chi_P^K(T)=&\int_T\diff x \left\langle{\cal T}K^a(x)K_a(0)\right\rangle=-\frac{\condl (T)+2\conds (T)}{m_l + m_s},
\label{WIK}\\
 \chi_S^\kappa(T)=&\int_T\diff x \left\langle{\cal T}\kappa^a(x)\kappa_a(0)\right\rangle=\frac{\condl (T)-2\conds (T)}{m_s-m_l},
\label{WIkappa}
\end{align}
where $\condl=\langle \bar u u + \bar d d\rangle$ and $\conds$ are the light- and strange-quark condensates, 
$\int_T dx\equiv \intT$ at a temperature $T\neq 0$,
\begin{equation}
  K^a=i\bar \psi\gamma^5\lambda^a \psi,\quad \kappa^a=\bar \psi\lambda^a \psi,\quad a=4,\cdots,7,\label{Kkbil}
\end{equation}
are the pseudoscalar and scalar $I=1/2$ quark bilinears, whose lightest states are  the kaon and $K_0^*(700)$ mesons, respectively, and $\psi$ is the quark triplet.

The $K$ and $\kappa$ bilinears in (4) can be related by both a chiral $O(4)$ and a $U(1)_A$ transformation~\cite{GomezNicola:2017bhm}. 
Namely, a general $SU(2)_A\times U(1)_A$ rotation of the up and down quark fields  
$$\displaystyle\psi'\to e^{i \gamma_5(\alpha_0\, \mathbb{1}_2 +\alpha_b\, \tilde\tau^b)}\psi,$$
with $\mathbb{1}_2=\text{diag}(1,1,0)$, $\tilde\tau^b=\left(\begin{array}{cc}\tau^b &\\ &0\end{array}\right)$, 
and $b=1,2,3$, acting on the $K^a$ bilinear
\begin{align}
K^a(x)'\rightarrow&\cos\alpha_0(x)\cos\alpha_b(x)K^a(x)\nonumber\\
                   -&\sin\alpha_0(x)\cos\alpha_b(x)\kappa^a(x)\nonumber\\
                   -&2d_{abc}\sin\alpha_b(x)\kappa^c(x),\nonumber\\
\text{with}\qquad  d_{abc}=&\pm 1/2, \quad a,c=4,\cdots,7,
\end{align}
connects it with $\kappa^a$ field. This connection has some important consequences:
\begin{enumerate}
\item Both $O(4)$ and $U(1)_A$ exact restoration imply $K-\kappa$ degeneration. 
\item The opposite is not necessarily true; there might be a region where $\chi_P^K\sim \chi_S^\kappa$ but the $O(4)$ or $U(1)_A$ symmetries are still significantly broken, i.e.,  $K-\kappa$ degeneration is a necessary but not sufficient condition for  $O(4)\times U(1)_A$ restoration
\footnote{The same caveat actually applies to most of the observables employed to study $O(4)$ and $U(1)_A$ restoration, like $\pi-\sigma$, $\pi-\delta$ degeneration or  $\condl\to 0$.}. 
Nevertheless, throughout this work we will provide evidences supporting the actual connection between $K-\kappa$ degeneration and $O(4)\times U(1)_A$  restoration.

\item Exact $O(4)$ restoration at $T_c$ takes place only for $N_f=2$ in the light chiral limit $m_l\to 0$.
Note that in this case, WIs for pure $U_A(1)$ observables, like the topological susceptibility or the difference between the $\pi$ and $\eta_l$ susceptibilities, imply $O(4)\times U(1)_A$ restoration at the exact $O(4)$ transition~\cite{GomezNicola:2017bhm,Nicola:2018vug}.
Thus, in this limit $K-\kappa$ should also degenerate at $T_c$.


\item In the physical case  with $N_f=2+1$ and nonzero quark masses, $\condl$ vanishes only asymptotically and the $O(4)$ symmetric phase is reached only approximately. In addition, the $U(1)_A$ symmetry vanishes also asymptotically above $T_c$. Thus, in this case $\chi_P^K-\chi_S^\kappa$ might be still different to zero above the crossover chiral transition. 

\item In the $SU(3)$ limit, i.e., for $m_s=m_l$, the $K$ and $K_0(700)/\kappa$ are expected to degenerate with the $\pi$ and $f_0(500)/\sigma$, respectively ~\cite{Oller:2003vf,RuizdeElvira:2017aet} and hence, one should expect $K-\kappa$ degeneration at the chiral transition, understood as the region where $\pi$ and $\sigma$ degenerate.

\end{enumerate}


Furthermore, recent theoretical analyses from the Nambu–Jona-Lasinio (NJL) model~\cite{Ishii:2016dln} and Chiral Perturbation Theory (ChPT)~\cite{Nicola:2018vug} have shown that in the physical case $K-\kappa$ degeneration occurs  above the crossover region, but around the same temperature where the isoscalar and isovector $O(4)\times U(1)_A$ partners degenerate. Actually, within ChPT, the temperature at which $\chi_P^K$ matches $\chi_S^\kappa$ is practically the same at which the $U(1)_A$ partners $\pi$ and $\delta$ degenerate. Consistent results pointing in the same direction are obtained from lattice analyses of $K$ and $\kappa$ screening masses~\cite{Cheng:2010fe,Bazavov:2019www}, which only degenerate at temperatures above 200 MeV, again in the same region of $O(4)\times U(1)_A$ partner degeneration. 

In that context, the advantage of the WIs~\eqref{WIK}-\eqref{WIkappa} is that $\chi_P^K$ and $\chi_S^\kappa$ are expressed in terms of well-measured quark condensates, whose thermal behavior provide a model independent tool to study their degeneration. 
Before discussing in more detail the consequences of $K-\kappa$ degeneration for $O(4)\times U(1)_A$ restoration, we will first analyze in the next section what can be learned about $\chi_P^K$ and $\chi_S^\kappa$ using the WIs \eqref{WIK}-\eqref{WIkappa}.


\section{Properties of $\chi_P^K$ and $\chi_S^\kappa$}\label{sec:chips}

\begin{figure*}
  \centering
  \includegraphics[width=0.9\textwidth]{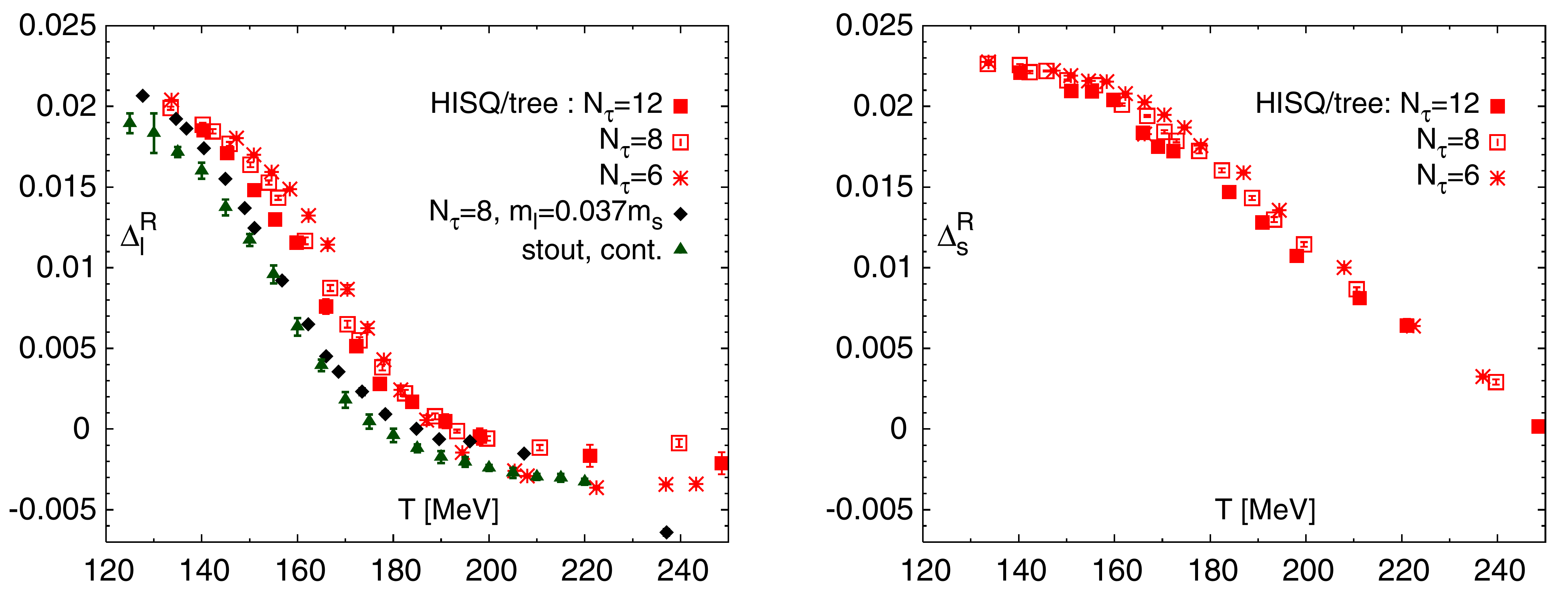}
  \caption{Light- (left panel) and strange-quark (right panel) subtracted  condensates around $T_c\sim 155$ MeV computed in~\cite{Borsanyi:2010bp,Bazavov:2011nk}.
They are defined as $\Delta^R_{l,s}=d+2m_sr_1^4 \left(\langle\bar \psi \psi\rangle_T-\langle\bar \psi \psi\rangle_0\right)$, with $\psi=q,s$ and $d$, $r_1$ lattice parameters defined in~\cite{Bazavov:2011nk}. Figure taken from~\cite{Bazavov:2011nk}.}\label{fig:conds}
\end{figure*}

The light-quark condensate $\condl$, as the order parameter of the chiral transition, is expected to drop abruptly at the transition temperature, with an inflection point at $T_c$ for physical quark masses. However, $\conds$ is supposed to decrease much softly due to the explicit chiral symmetry breaking of the heavier strange quark~\cite{Bazavov:2011nk}, being $m_l/m_s$ the parameter regulating the relative drop of these two condensates. These  trends can be clearly observed in lattice analyses. For illustrative purposes we show in Fig.~\ref{fig:conds}  the  light ($\Delta^R_l$) and strange ($\Delta^R_s$) subtracted condensates reported in~\cite{Bazavov:2011nk}; while $\Delta^R_l$ drops abruptly close to the chiral transition and asymptotically above $T_c$, $\Delta^R_s$ remains large at the critical temperature, showing only a smooth decrease. Note that lattice quark condensates usually have to be subtracted to remove UV divergences $\langle \bar q_i q_i \rangle \sim m_i/a$, with $a$ the lattice spacing and $m_i$ the quark mass as defined in~\cite{Bazavov:2011nk}. 

On the one hand, since both the light- and strange-quark condensates are negative quantities,~\eqref{WIK} indicates that $\chi_P^K$ should decrease continuously at all temperatures, with an abrupt drop off around $T_c$ coming from the light-quark condensate.
On the other hand,~\eqref{WIkappa} implies that below and around the chiral transition $\chi_S^\kappa(T)$  should grow following the $\condl$ decrease and the roughly constant behavior of $\conds$. Nevertheless, above the $O(4)$ transition the light-quark condensate starts decreasing only asymptotically, while at some temperature the $\conds$ reduction takes over, hence changing the trend of $\chi_S^\kappa(T)$ to a  slowly decreasing behavior towards degeneration with $\chi_P^K (T)$.

The previous argument implies that in the physical case, $\chi_S^\kappa$ should have a maximum at a temperature $T>T_c$ and that the behavior of the curve above the maximum 
 is driven by the $\conds$ drop. In addition, near the chiral $m_l/m_s\rightarrow 0$ limit one should expect a steepest growth below the maximum, dictated by $\condl$, but a flattening above it, from $\conds$, pointing out for $K-\kappa$ degeneration 
at lower temperatures.  On the contrary, in the $SU(3)$ $m_l/m_s\rightarrow 1$ limit, the peak should be more pronounced from both sides, consistently with $\kappa$ and $\sigma$ degeneration. 

The existence and properties of the $\chi_S^\kappa (T)$ peak, coming from WIs and confirmed with our lattice and theoretical UChPT analysis  below, are key results of the present work.

\subsection{Results from lattice data}\label{sec:latt}

\begin{figure}
  \includegraphics[width=0.495\textwidth]{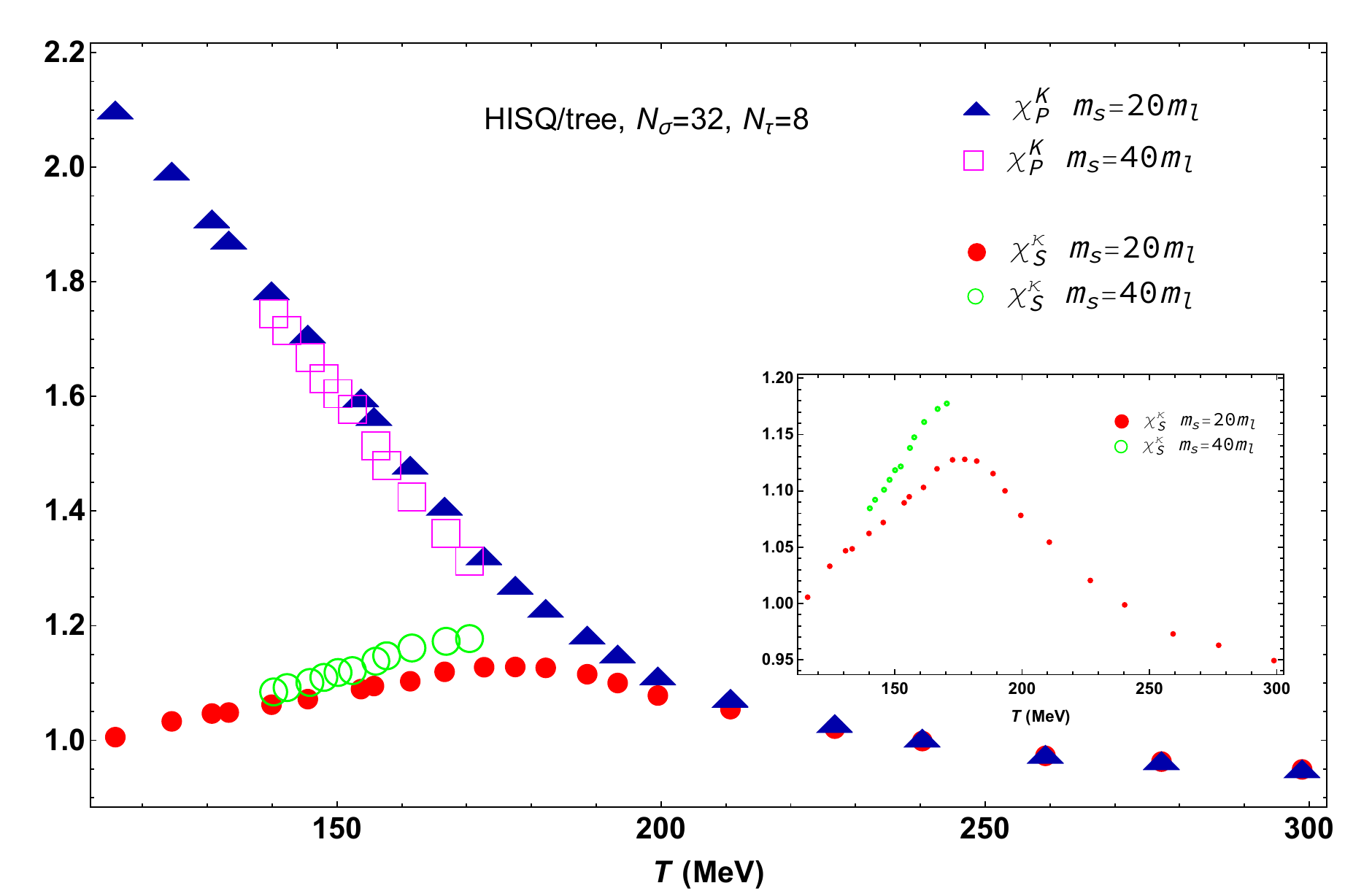}
  \caption{Reconstructed pseudoscalar and scalar susceptibilities (in lattice units) in the $I=1/2$ channel from~\eqref{WIK}~and~\eqref{WIkappa}, respectively, with 
    light- and strange-quark condensate data from~\cite{Bazavov:2011nk,Bazavov:2014pvz}. In the inset panel we show $\chi_S^\kappa (T)$ separately in order to emphasize the peak behavior.   For such lattice setup, the continuum extrapolation to the physical mass case $m_s=27\,m_l$  gives $T_c=154\pm 9$ MeV.}
   \label{fig:chis}
\end{figure}

Without direct $\chi_S^\kappa(T)$ and  $\chi_P^K(T)$ lattice data available, the above hypotheses can be tested using lattice results for the combinations of light- and strange-quark condensates appearing in~\eqref{WIK} and~\eqref{WIkappa}, which we denote as reconstructed susceptibilities. In Fig.~\ref{fig:chis} we show the results of the reconstructed susceptibilities using the unsubtracted condensate data in~\cite{Bazavov:2011nk,Bazavov:2014pvz} for two different quark-mass configurations: $m_s=20\,m_l$, which is close to the physical point, and $m_s=40\,m_l$, closer to the chiral limit. 

The results in Fig.~\ref{fig:chis} are fully consistent with the theoretical expectations. First, $\chi_P^K$ decreases at all temperatures, with a smooth asymptotic behavior above  $T_c$. Second, the scalar susceptibility $\chi_S^\kappa$ develops a maximum around $T_c$, after which it shows  a smooth temperature dependence,  degenerating with $\chi_P^K$ at temperatures  $T\,\gsim 180$ MeV. 
Such degeneration  is ultimately driven by strangeness through $\conds$.
In fact, the data for $m_s=40\,m_l$ reflect the expected behavior below $T_c$  and seem to indicate $\chi_S^\kappa-\chi_P^K$  degeneration at lower temperatures, as expected from the softer $\conds$ $T$-dependence. Unfortunately, available lattice results for condensates closer to the chiral limit~\cite{Bazavov:2011nk,Ding:2019prx} do not reach the temperatures of interest for our purposes here.  

The results in Fig.~\ref{fig:chis} are also supported by lattice analyses of screening masses~\cite{Cheng:2010fe,Bazavov:2019www}.
Although screening masses and susceptibilities measure different limits of meson correlators at vanishing four momenta (the susceptibility behaves as the inverse pole mass squared while the screening mass measures the exponential falloff of the correlator at large spatial separation), these two quantities follow a similar temperature scaling~\cite{Nicola:2016jlj,Nicola:2018vug,Ishii:2016dln}.
Lattice results show a minimum around the transition for the $\kappa$ screening mass, hence playing the counterpart of the $\chi_S^\kappa$  maximum in Fig.~\ref{fig:chis}.
Moreover, all scalar channels analyzed in~\cite{Cheng:2010fe,Bazavov:2019www} show a similar minimum driven by its degeneration with their corresponding pseudoscalar partner, which might constitute a global meson pattern. 

\subsection{Theoretical analysis from Effective Theories}\label{sec:uchpt}

In order to analyze the behavior of $\chi_S^\kappa$ from the theoretical side at and  beyond the physical point, 
we consider a UChPT approach where $\chi_S^\kappa (T)$ is saturated by its lowest pole, the $K_0^*(700)/\kappa$ meson, generated in $\pi K$ scattering at finite temperature. Namely,
\begin{equation}
\chi_S^{\kappa, U}(T)=A_\kappa \frac{M_\kappa^2 (0)}{M_\kappa^2 (T)}
\label{chisat}
\end{equation}
where we fix $A_\kappa$  to reproduce the perturbative ChPT result at $T=0$, i.e., $A_\kappa=\chi_S^{\kappa,\text{ChPT}} (0)$ calculated in~\cite{Nicola:2018vug},  and $M_\kappa^2 (T)=M_p^2(T)-\Gamma_p^2(T)/4$,
with $s_p=(M_p-i\Gamma_p/2)^2$ the resonance pole position in the second Riemann sheet of the complex $s$-plane for the unitarized $\pi K$ $I=1/2$ scalar partial wave.
Thus, $M_\kappa^2(T)$ is the real part of the  $K_0^*(700)$ self-energy at the pole, which is expected to provide the dominant temperature dependence of $\chi_S^\kappa$.
Since susceptibilities are $p=0$ correlators, the sensibility to the $p$-dependence of the self-energy and the $T$-dependence of its residue in~\eqref{chisat} are assumed to lie within the uncertainty bands.
That is actually the case when the scalar susceptibility $\chi_S(T)$ is saturated by the thermal $f_0(500)/\sigma$~\cite{Nicola:2013vma,Ferreres-Sole:2018djq}. Namely, this  approach  
has been proven to reproduce the $\chi_S (T)$ transition peak and to describe lattice data around it.

For the unitarized $\pi K$ scattering amplitude we rely on  the UChPT techniques described in~\cite{Oller:1997ti,Ferreres-Sole:2018djq,Gao:2019idb} and write the $I=1/2$ scalar $\pi K$ partial wave as
\begin{equation}
t_{U}(s;T)=\frac{t_2^2(s)}{t_2(s)-\tilde t_{4}(s,T)},
\label{tunitmod}
\end{equation}
where $s=(p_\pi+p_K)^2$,
$$t_2(s)=\frac{5s^2-2s(M_K^2+M_\pi^2)-3(M_K^2-M_\pi^2)^2}{128F_\pi^2}$$ 
is the $T$-independent leading-order $\Od(p^2)$ ChPT amplitude, $M_{\pi (K)}$ the pion (kaon) mass and  $F_\pi$ the pion decay constant.
For the $\Od(p^4)$ contribution $\tilde t_{4}(s,T)$ we consider two different methods consistent within unitarity and analyticity requirements for the thermal amplitude:
\begin{eqnarray}
\mbox {Method 1:} \quad  \tilde t_4(s;T)&=&16\pi\, t_2(s)^2 \tilde J_{\pi K} (s;T),\nonumber   \\
\mbox {Method 2:} \quad  \tilde t_4(s;T)&=& t_4(s;0)+16\pi\, t_2(s)^2   \left[J_{\pi K} (s;T), \right.\nonumber\\
&-& \left.J_{\pi K} (s;0)   \right]  \nonumber
\end{eqnarray}
where 
\begin{eqnarray}
J_{\pi K} (s;T) &=& T\sum_{n=-\infty}^\infty\qint \frac{1}{q^2-M_K^2}\frac{1}{(q-Q)^2-M_\pi^2}\nonumber \\
\label{Jther}
\end{eqnarray}
is the one-loop thermal integral in the center-of-momentum frame, whose detailed expression can be found, e.g., in~\cite{Nicola:2014eda},
with $q_0=2\pi i\,n\,T$, $\vec{Q}=\vec{0}$ and $Q_0^2\rightarrow s$ after analytic continuation from external discrete frequencies.

Method 1 was proposed in~\cite{Gao:2019idb}, where $\tilde J_{\pi K}$ denotes the finite part of $J_{\pi K}$,  renormalized by a subtraction constant fitted to scattering data at $T=0$~\cite{Buettiker:2003pp,Ledwig:2014cla}. 
In Method 2, $t_4(s;0)$ is the renormalized ChPT $\Od(p^4)$ amplitude at $T=0$~\cite{GomezNicola:2001as}.
The main advantage of Method 2 is that is consistent with the perturbative chiral expansion at $\Od(p^4)$ at $T=0$; hence, providing better control over the quark mass dependence of the amplitude.
Similarly to the $\pi\pi$ scattering case studied in~\cite{GomezNicola:2002tn,Dobado:2002xf}, both methods ensure elastic thermal unitarity~\cite{Nicola:2014eda}, which for $\pi K$ scattering reads:
\begin{equation}
\im t_U(s;T)=\sigma_{\pi K}(s;T)\vert  t_U(s;T) \vert^2,\quad s\geq (M_K+M_\pi)^2, \label{therunit}
\end{equation}
where the thermal phase-space factor 
\begin{eqnarray}
\sigma_{\pi K}(s;T)&=&\frac{1}{s}\sqrt{\left(s-(M_\pi+M_K)^2\right)\left(s-(M_\pi-M_K)^2\right)}\nonumber\\
&\times& \left[  1+ n(E_+)+n(E_-) \right], 
\end{eqnarray}
with $E_{\pm}=(s\pm\Delta)/(2\sqrt{s})$, $\Delta=M_K^2-M_\pi^2$ and $n(x)=(e^{x/T}-1)^{-1}$ is the Bose-Einstein distribution function.  

As a  test of the capability of Methods 1 and 2 to describe the $K_0^*(700)$, we get at $T=0$  $\sqrt{s_p}^{(1)}=(731\pm 7)-i(280\pm 9)$ MeV and $\sqrt{s_p}^{(2)}=(679\pm 6)-i(289\pm 8)$ MeV, for method 1 and 2, respectively, where we have used the subtraction constant value and error in~\cite{Gao:2019idb} for method 1 and the Low-Energy Constants (LECs) of the global fit in~\cite{Molina:2020qpw} for method 2. For the latter, the uncertainties are computed from the propagation in quadrature of the LEC errors.
These results are perfectly consistent with the most precise dispersive calculations~\cite{DescotesGenon:2006uk,Pelaez:2016klv,Pelaez:2020uiw}.
\begin{figure*}
  \includegraphics[width=0.465\textwidth]{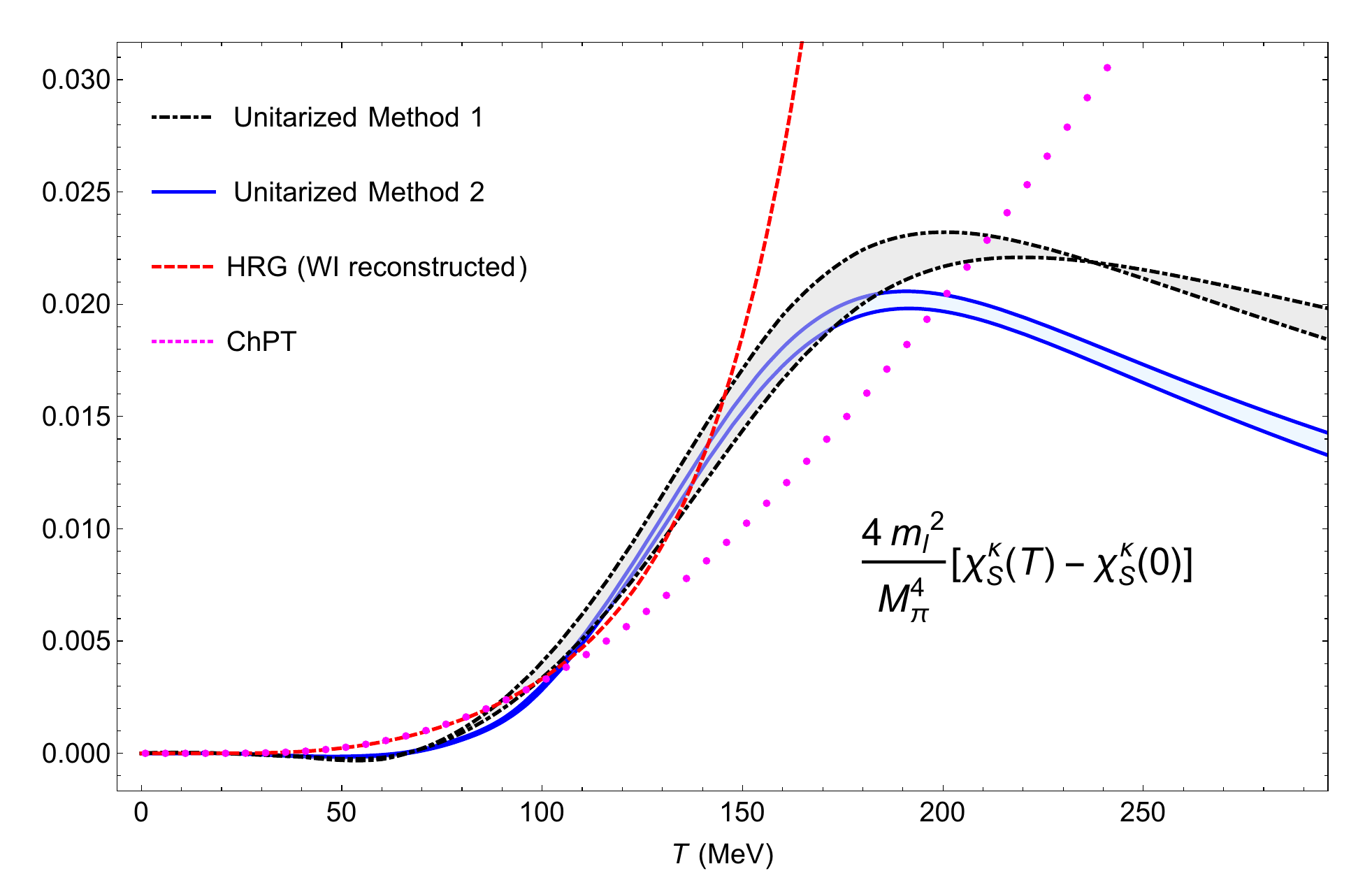}
   \includegraphics[width=0.45\textwidth]{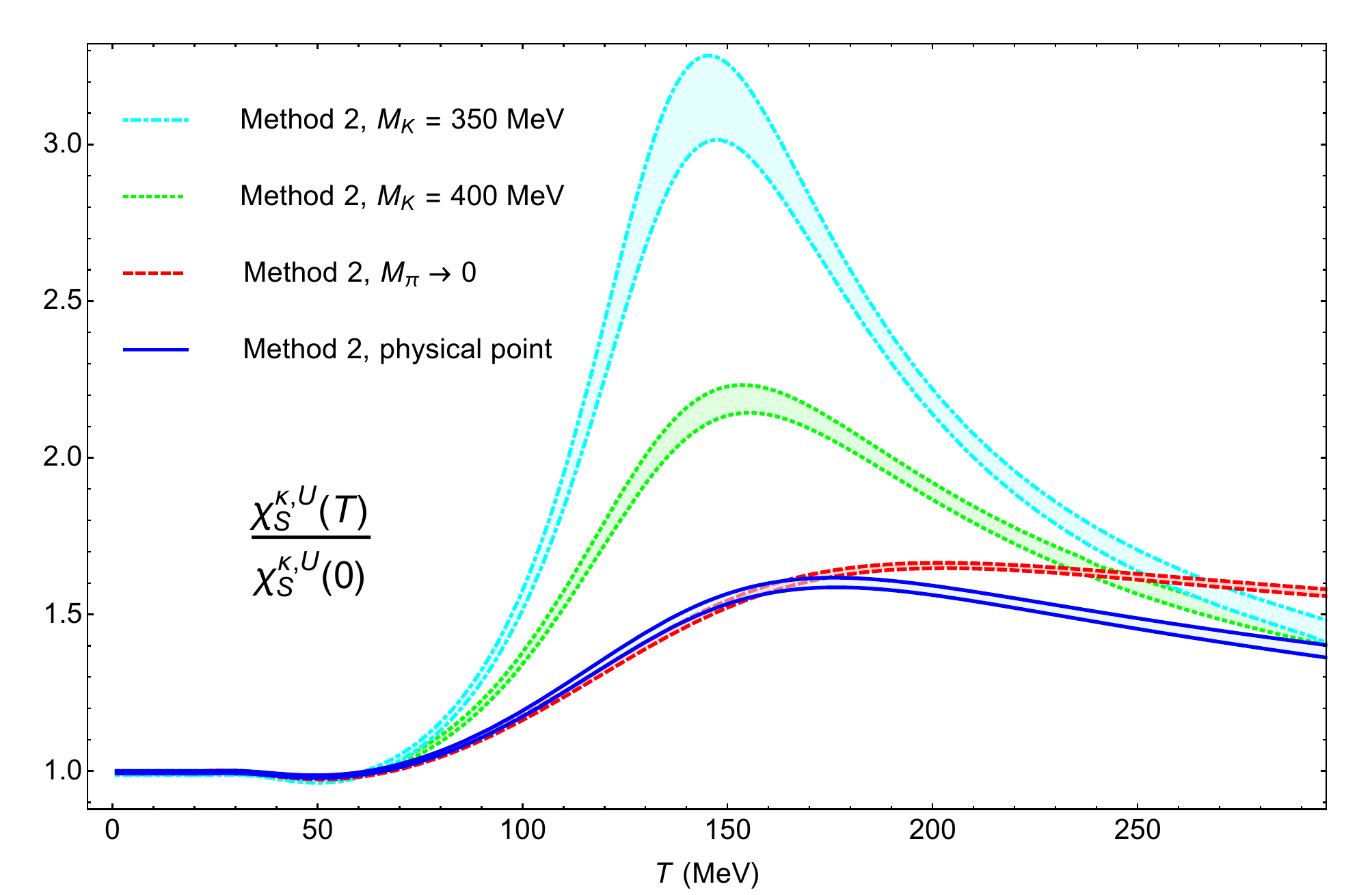}
  \caption{Left panel:  $\kappa$ susceptibility from UChPT with the two methods explained in the main text, including their theoretical uncertainty bands,  the perturbative ChPT result and the HRG  one extracted from the quark condensate combination in \eqref{WIkappa}. 
  Right panel: Unitarized $\kappa$ susceptibility for physical masses, the light chiral limit and for $M_K$ closer to the $SU(3)$ limit.}
   \label{fig:susunit}
\end{figure*}

The $\chi_S^{\kappa,U}$  results for both unitarization methods, including their uncertainties, are plotted in Fig.~\ref{fig:susunit} together with 
the reconstructed scalar susceptibility from the WI in~\eqref{WIkappa} using for the light- and strange-quark condensates
the Hadron Resonance Gas (HRG) results in~\cite{Jankowski:2012ms}. The HRG includes hadron resonances of masses below 2 GeV and is meant to capture the relevant thermodynamics below the transition, hence providing a check of consistency for our unitarized results and a way to estimate the importance of such higher states for this observable.  In addition, we  also include the perturbative ChPT prediction calculated in~\cite{Nicola:2018vug}. The two thermal unitarized methods remain fairly consistent between them and compatible with the perturbative ChPT and HRG results below the transition, indicating the robustness of the approach.  In addition, both reproduce the expected $\chi_S^{\kappa,U}$ peak behavior, unlike the ChPT or HRG, which are  monotonically increasing. This reveals the importance of considering thermal interactions in order to describe these non-perturbative phenomena around the transition. 


 Finally, we study the chiral and $SU(3)$ limits using method 2, which at $T=0$ reproduces the light- and strange-quark mass dependence of the $\pi K$ amplitude predicted in ChPT.
 The behavior of $\chi_S^{\kappa,U}$ in the two limits is plotted in Fig.~\ref{fig:susunit}, showing the expected results. On the one hand, when approaching the chiral limit we find a steepest growth below the maximum and a flatter curve above it.
 On the other hand, the reduction of the kaon mass enhances the size of the peak and moves it to lower temperatures closer to $T_c$, so that  $\chi_S^{\kappa,U}$ tends to resemble the behavior of $\chi_S$ consistently with $SU(3)$ symmetry.
 
 The above results for the $I=1/2$ scalar susceptibility within the UChPT approach constitute also an important outcome of the present work. 

\section{Consequences for $O(4)\times U(1)_A$ restoration}
\label{sec:conseq}

The results described in Section~\ref{sec:chips} show that $\chi_S^\kappa$ develops a maximum, after which it degenerates with $\chi_P^K$.
This prediction is obtained from rigorously derived WIs and hence, it can be considered as a model independent result. In that context, some comments are in order to relate this sector with $O(4)\times U(1)_A$ restoration:
\newline

(i)- Close to the physical point, i.e., using the $m_s=20\, m_l$ lattice data for the light- and strange-quark condensates in~\cite{Bazavov:2011nk,Bazavov:2014pvz},
one observes that the position of the $\chi_S^\kappa$ peak lies well above the $O(4)$ crossover region---the temperature at which $\chi_S$ develops a maximum and $\pi-\sigma$ degenerate. There is no contradiction since $K-\kappa$ degeneration  should happen at $T_c$ only if $O(4)$ restoration is exact i.e., for $N_f=2$ and $m_l\to 0$.

%

Conversely, the temperature at which the reconstructed $\chi_S^\kappa$ and $\chi_P^K$ degenerate lies close to region where current $N_f=2+1$ lattice data~\cite{Buchoff:2013nra,Dick:2015twa} find $\pi-\delta$ degeneration; even when $K-\kappa$ degeneration only imposes a lower bound for $U(1)_A$ restoration, this result shows that, for physical quark masses, such degeneration lies around the point where $U(1)_A$ symmetry is assumed to be asymptotically restored in $N_f=2+1$ lattice simulations. 
Thus, $K-\kappa$ degeneration might be considered as an additional sign to study asymptotic $O(4)\times U(1)_A$ restoration.
Additional arguments supporting this proposal are the degeneration of $K-\kappa$ lattice screening masses in the $T\sim$ 200 MeV region of $O(4)\times U(1)_A$ restoration~\cite{Bazavov:2019www}, and results in ChPT and NJL models showing coincidence of the $K-\kappa$ degeneration temperature with that of the $U(1)_A$ partners, like $\pi-\delta$~\cite{Ishii:2016dln,Nicola:2018vug}. 

\begin{figure}
  \centering
  \includegraphics[width=0.49\textwidth]{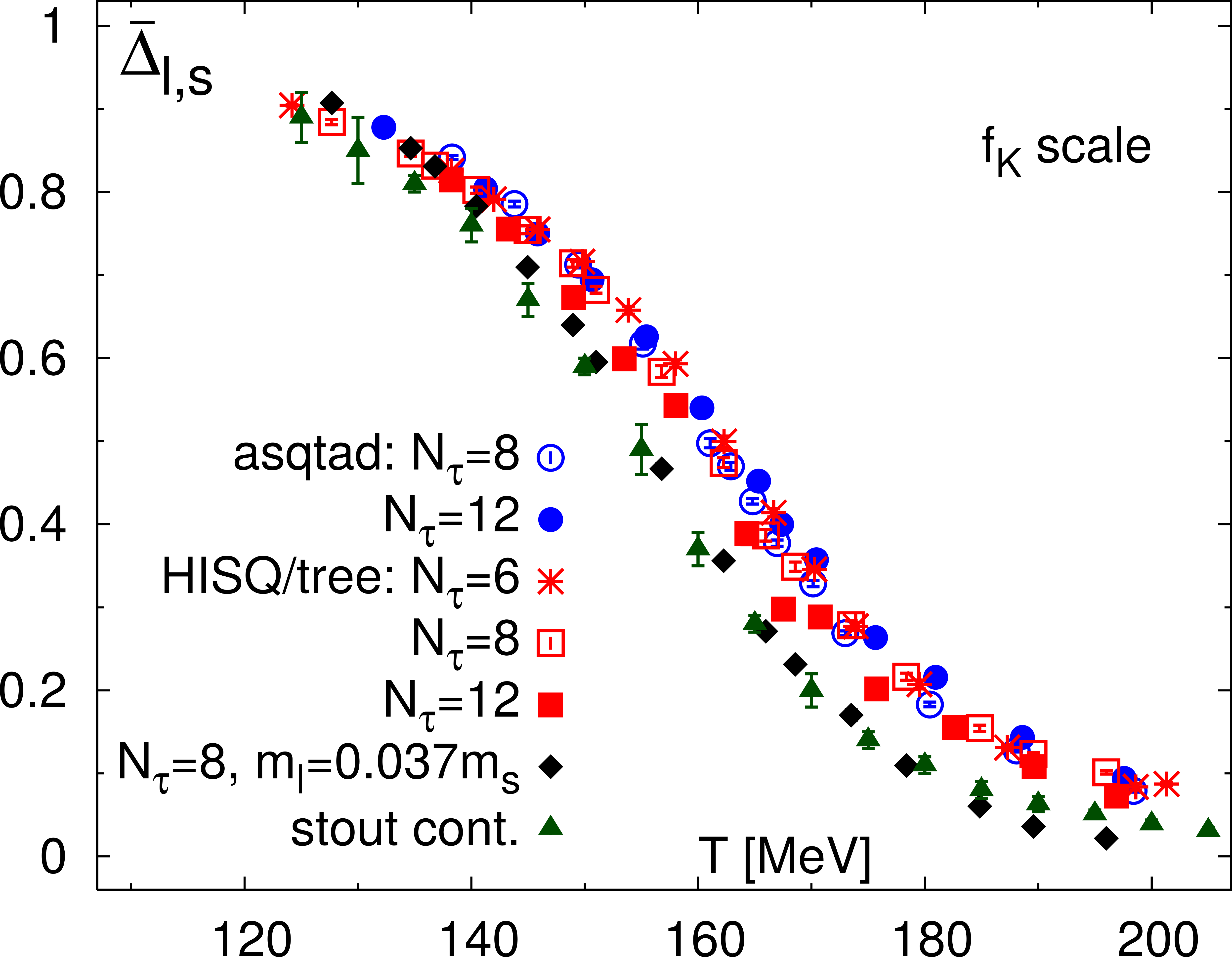}
  \caption{Subtracted quark condensate  normalized to its value at zero temperature as given in~\cite{Bazavov:2011nk}. At $T_c$, $\bar\Delta_{l,s}$ has reduced its value by half and only at higher temperatures it shows an asymptotic vanishing behavior compatible with $O(4)\times U(1)_A$ restoration. Figure taken from~\cite{Bazavov:2011nk}.}\label{fig:deltals}
\end{figure}

Further information can be obtained by taking the difference between~\eqref{WIK} and~\eqref{WIkappa},
\begin{equation}
  \chi_S^\kappa (T)-\chi_P^K (T)=\frac{2}{m_s^2-m_l^2}\,\Delta_{l,s}(T),
  \label{WIdif}
\end{equation}
where $\Delta_{l,s}(T)=m_s\condl(T)-2m_l \conds(T)$ is the so-called subtracted condensate, one of the order parameters considered in the lattice literature\footnote{The renormalized subtracted condensates plotted in Fig.~\ref{fig:conds} are another commonly employed choice.} to cancel out finite-size divergences~\cite{Aoki:2009sc,Borsanyi:2010bp,Bazavov:2011nk,Ratti:2018ksb}. Eq.~\eqref{WIdif} provides information on  
$K-\kappa$ degeneration in terms of a well-determined lattice quantity. In Fig.~\ref{fig:deltals} we plot the normalized $\bar\Delta_{l,s}(T)=\Delta_{l,s}(T)/\Delta_{l,s}(0)$ results given in~\cite{Bazavov:2011nk}, where  one can see that at $T_c$ this difference  has reduced its value by half, and only at much larger temperatures $T> 200$ MeV $\Delta_{l,s}$ shows an asymptotic vanishing behavior compatible with effective $O(4)\times  U(1)_A$ restoration. Thus, even when $\Delta_{l,s}$ is usually considered equivalent to $\condl$, i.e., as an order parameter for $O(4)$ restoration, this is only true in the light chiral limit. In the physical case,  Eq.~\eqref{WIdif} indicates that its vanishing actually provides a sign of $O(4)\times U(1)_A$ restoration. Note, however, that its inflection point practically coincides with that of $\condl$ since the variation of $\conds$ with $T$  is almost negligible at that temperatures and hence, the $\Delta_{l,s}$ subtracted condensate works perfectly well to estimate the crossover temperature. This is  not in conflict with the $\Delta_{l,s}$ vanishing signaling $O(4)\times U(1)_A$ restoration, being $\conds$ ultimately responsible for the $\Delta_{l,s}$ tail behavior.
\newline 

(ii)- Let us consider now the $m_s\gg m_l$ expansion. On the one hand, the right-hand side in~\eqref{WIdif} reads
\begin{equation}
  \chi_S^\kappa (T)-\chi_P^K (T)\big\vert_{m_s\gg m_l}=\frac{2}{m_s}\,\condl(T)\Big\vert_{\text{SU(2)}}+ {\cal O}\left(1/m_s^{2}\right).
  \label{WIdifnf2}
\end{equation}

On the other hand, the left-hand side of~\eqref{WIdif} in this limit can  be studied in ChPT, where at leading order in this expansion one finds  $\chi_S^\kappa (T)-\chi_P^K (T)\big\vert_{m_s\gg m_l}\propto 1/m_s$~\cite{Nicola:2018vug}.   Thus, at leading order, this regime is nothing but the $N_f=2$ limit, where the strange quark is fully decoupled for $m_s\to\infty$. Even though the $K$ and $\kappa$ susceptibilities are pure $SU(3)$ quantities and, in the physical case, the strange quark should be taken as a dynamical degree of freedom, its difference has a well-defined $N_f=2$ limit.

Furthermore, Eq.~\eqref{WIdifnf2} implies $K-\kappa$ degeneration at $T_c$ in the $m_l\to0$ limit, since in this case the light-quark condensate vanishes exactly.
This is consistent with the analysis in~\cite{GomezNicola:2017bhm,Nicola:2018vug} and two-flavor lattice results~\cite{Aoki:2012yj,Cossu:2013uua,Tomiya:2016jwr,Brandt:2016daq,Brandt:2019ksy}, which suggest $O(4)\times U(1)_A$ restoration at the $O(4)$ transition for $N_f=2$ in the light chiral limit. Thus, our analysis helps then to reconcile lattice results in these two different regimes. 
\newline

(iii)- It is also relevant to discuss the light chiral limit for $N_f=2+1$.  As can check in ChPT, both $\chi_S^\kappa$ and $\chi_P^K$ are well behaved quantities in this case~\cite{Nicola:2018vug}, and hence, Eq.~\eqref{WIdif} simplifies to
\begin{equation}
  \chi_S^\kappa (T)-\chi_P^K (T)\big\vert_{m_l=0}=\frac{2}{m_s}\,\condl(T)\Big\vert_{m_l=0}.
  \label{WIdifch}
\end{equation}

Note that the difference between~\eqref{WIdifnf2} and~\eqref{WIdifch} is that for the latter, the light-quark condensate appearing in the right-hand side is the $SU(3)$ result in the light chiral limit, which does not vanish at $T_c$ and hence, the $K-\kappa$ susceptibility difference does not vanish either. Even though the susceptibility difference in this case is expressed in terms of the chiral condensate\footnote{This comes as no surprise since exact $O(4)$ implies $K-\kappa$ degeneration.}, Eq.~\eqref{WIdifch} does not imply any consequence regarding $O(4)\times U(1)_A$ restoration since both symmetries are explicitly broken.
Clearly, $m_l\to 0$ accelerates chiral $O(4)$ restoration with respect to the physical case, but it also does so with $U(1)_A$ restoration, since we are closer to the regime where exact $O(4)$ restoration implies an exact $O(4)\times U(1)_A$ symmetric phase. This is actually reflected in the behavior of the $\chi_S^\kappa$ peak in the light chiral limit analyzed in section~\ref{sec:chips} both for the reconstructed lattice data and the UChPT. Only when the  $m_l/m_s\rightarrow 0$ limit is taken, Eqs.~\eqref{WIdifch}~and~\eqref{WIdifnf2} coincide, the strange quark decouples and the previous conclusion about $O(4)\times U(1)_A$ restoration at $T_c$ follows. 
 \newline

%


 (iv)- The $SU(3)$ $m_l/m_s \to 1$ limit in \eqref{WIdif} is also well behaved and illustrative. Evaluating it once again in ChPT, we get
\begin{equation}
  \chi_S^\kappa (T)-\chi_P^K (T)\big\vert_{m_s\to m_l}=\chi_S+\frac{\condl}{m_l}=\chi_S-\chi_P^\pi,
  \label{WIdifsu3}
\end{equation}
where $\chi_P^\pi$ stands for the pion pseudoscalar susceptibility. This confirms our previous statement about this limit, where the $\kappa$ and the $I=0$ scalar susceptibilities degenerate one into another. Thus, if this limit is obtained by reducing the strange-quark mass but keeping the light-quark mass fixed, one expects $\chi_S^\kappa$ to resemble the $\chi_S$ crossover peak, as we do observe in our UChPT analysis.
Finally, it is worth noting that even when the $K-\kappa$ susceptibility difference is expressed once more in terms of two $O(4)$ quantities, they are nonzero at $T_c$ in $N_f=2+1$. Thus, $O(4)\times U(1)_A$ would not be restored near $T_c$ in this limit because one is far away from the strange-quark decoupling regime. 

Our arguments in this section justify then that $m_l/m_s$ is the relevant parameter regarding the behavior of the $K-\kappa$ susceptibility difference and its connection with $O(4)\times U(1)_A$ restoration.

\section{Conclusions}

We have performed a detailed analysis of the scalar and pseudoscalar susceptibilities in the $I=1/2$ channel based on Ward Identities, lattice data and Unitarized Chiral Perturbation Theory, which provides alternative way to study the interplay between chiral and $O(4)\times U(1)_A$ restoration and the role of strangeness in that context.
This joint analysis gives rise to the following consistent results:
\newline


1)- The $\kappa$ scalar susceptibility develops a peak, which in the physical limit and for $N_f=2+1$ lies above the chiral crossover. Below this peak, the rise of the susceptibility is controlled by the light-quark condensate and then it is mostly related to chiral restoration. Above the peak, the susceptibility drop is driven by the strange-quark condensate and the $\kappa$ susceptibility tends to degenerate with the $K$ one. 
\newline

2)- Although there are no direct lattice results for the $I=1/2$ susceptibilities available, we reconstruct them from WIs and condensate data. The results confirm the existence of the $\chi_S^\kappa$ peak and the $\chi_S^\kappa-\chi_P^K$ degeneration. In the physical case, the position of the peak lies within the region of $O(4)\times U(1)_A$ restoration, i.e., the temperature where lattice analyses suggest the vanishing of the topological susceptibility or $\pi-\delta$ degeneration. 
\newline

3)- Within a UChPT approach, we have studied  the $\kappa$ scalar susceptibility by saturating it with the thermal pole of the $K_0^* (700)/\kappa$, the lightest $I=1/2$ scalar state, which is  dynamically  generated through unitarized $\pi-K$ scattering at finite temperature. The result confirms again the presence of the peak, which other approaches such as ChPT or the HRG are not able to provide, hence highlighting the importance of including properly thermal interactions.
Our analytic UChPT approach has also the advantage of allowing us to tune the meson masses beyond the physical limit to study the behavior of the $\kappa$ susceptibility in the chiral and $SU(3)$ limits. 
\newline

4)- The parameter $m_l/m_s$ controls effectively the transition from the $N_f=2$ to the $N_f=2+1$ cases for the observables analyzed here. This  offers a way to reconcile lattice results in these two scenarios regarding $O(4)\times U(1)_A$ restoration and suggest that $\chi_S^\kappa-\chi_P^K$ is as an alternative useful sign to study this problem.
In the physical case, our work, based on lattice data, is consistent with previous ChPT and NJL analyses, as well as with results from lattice screening masses, all pointing to $\chi_S^\kappa-\chi_P^K$ degeneration in the region where  $O(4)\times U(1)_A$ restoration takes place.

5)-The $\chi_S^\kappa-\chi_P^K$  susceptibility difference can be related to $\Delta_{l,s}$, one of the subtracted quark condensates customarily analyzed in the lattice.
For $m_l/m_s\to 0$, the strange quark decouples, the $N_f=2$ limit is reached and our analysis suggests that the $O(4)$ and $O(4)\times U(1)_A$ transitions coincide at $T_c$, hence consistently with previous WIs analyses and $N_f=2$ lattice data.
In the same way, our UChPT result shows a flattening of $\chi_S^\kappa$ above the peak in the light chiral limit, reflecting degeneration with $\chi_P^K$.
In the opposite limit, $m_l/m_s\to 1$, the degenerate $SU(3)$ phase is achieved, which implies $\chi_S^\kappa$ and $\chi_S$ degeneration. Within UChPT, we confirm this behavior by lowering the kaon mass, which makes the $\chi_S^\kappa$ peak grow and displace to the left towards $T_c$, thus resembling the behavior of $\chi_S$ at the crossover region. 


We believe that our present analysis provides new insight about the $I=1/2$ sector, which may be useful for future theoretical and lattice analyses. Furthermore, it helps to better understand the role of strangeness in the current tension between $N_f=2+1$ and $N_f=2$ lattice results regarding $O(4)\times U(1)_A$ restoration.

 
 \begin{acknowledgments}
 We are very grateful to  Z.~H.~Guo for useful comments and to F. Karsch and A. Lahiri for providing detailed lattice results.
 Work partially supported by research contract  PID2019-106080GB-C21 (spanish ``Ministerio de Ciencia e Innovaci\'on"),  the European Union Horizon 2020 research and innovation program under grant agreement No 824093 and the Swiss National Science Foundation, project No.\ PZ00P2\_174228. A. V-R acknowledges support from a fellowship of the UCM predoctoral program.
 \end{acknowledgments}

\end{document}